\documentclass[usegraphicx,usenatbib,useapjfonts,apj]{emulateapj}

\begin{document}

\def\gsim{\;\rlap{\lower 2.5pt
 \hbox{$\sim$}}\raise 1.5pt\hbox{$>$}\;}
\def\lsim{\;\rlap{\lower 2.5pt
   \hbox{$\sim$}}\raise 1.5pt\hbox{$<$}\;}
\newcommand{\vt}{\mbox{\bf {T}}}
\newcommand{\vtcmb}{\mbox{\bf {T}}}
\newcommand{\vm}{\mbox{\bf {M}}}
\newcommand{\vn}{\mbox{\bf {n}}}
\newcommand{\vv}{\mbox{\bf {v}}}
\newcommand{\vnh}{\hat {\mbox{\bf n}}}
\newcommand{\vf}{\mbox{\bf {F}}}
\newcommand{\vx}{\mbox{\bf {x}}}
\newcommand{\vy}{\mbox{\bf {y}}}
\newcommand{\vrv}{\mbox{\bf {r}}}
\newcommand{\vk}{\mbox{\bf {k}}}
\newcommand{\vq}{\mbox{\bf {q}}}
\newcommand{\vl}{\mbox{\bf {l}}}
\newcommand{\tcl}{\tilde{C}_l}

\def\degr{\hbox{$^\circ$}}
\input epsf 
\def\plotone#1{\centering \leavevmode \epsfxsize=\columnwidth \epsfbox{#1}}
\def\plotancho#1{\includegraphics[width=17cm]{#1}}
\title{Oxygen pumping II: Probing the Inhomogeneous Metal Enrichment at the
Epoch of Reionization with High Frequency CMB Observations}

\author{Carlos  Hern\'andez--Monteagudo\altaffilmark{1}, Zolt\'an Haiman\altaffilmark{2}, Licia Verde,\altaffilmark{3,4} and  Raul Jimenez\altaffilmark{3,4}
}

\altaffiltext{1}{Department of Physics and Astronomy, University of Pennsylvania, Philadelphia, PA 19104, USA; carloshm@astro.upenn.edu}
\altaffiltext{2}{Department of Astronomy, Columbia University, 550 West 120th Street, New York, NY 10027, USA; haiman@astro.columbia.edu}
\altaffiltext{3}{Institute of Space Sciences (CSIC-IEEC)/ICREA, Campus UAB, Bellaterra  
08193, Spain}
\altaffiltext{4}{Department of Astrophysical Sciences, Peyton Hall, Princeton University, Princeton NJ-08544, USA; raulj,lverde@astro.princeton.edu}


\begin{abstract}
  At the epoch of reionization, when the high-redshift inter-galactic
  medium (IGM) is being enriched with metals, the $63.2 \mu$m fine
  structure line of OI is pumped by the $\sim 1300$\AA\ soft UV
  background and introduces a spectral distortion in the Cosmic
  Microwave Background (CMB). Here we use a toy model for the spatial
  distribution of neutral oxygen, assuming metal bubbles surround dark
  matter halos, and compute the fluctuations of this distortion, and
  the angular power spectrum it imprints on the CMB. We discuss the
  dependence of the power spectrum on the velocity of the winds
  polluting the IGM with metals, the minimum mass of the halos
  producing these winds, and on the cosmic epoch when the OI pumping
  occurs. We find that, although the clustering signal of the CMB
  distortion is weak ($(\delta y)_{rms} \lsim  10^{-7}$) (roughly corresponding to a temperature anisotropy
  of $\sim$ nK), it may be reachable in deep integrations with high-sensitivity infrared detectors.
  Even without a detection, these instruments should be able to useful
  constraints on the heavy element enrichment history of the IGM.
  
\end{abstract}


\section{Introduction}
\label{sec:intro}

Probing the Dark Ages -- the epoch between the last scattering surface
of the CMB at $z\sim 1000$ and the completion of the reionization of
the IGM at $z\sim 6$, including the formation of the first luminous
objects -- constitutes the next frontier of observational cosmology.
Among the major open questions are the nature of the objects that
reionized the universe, and the origin of the first heavy elements, as
well as the efficiency with which they were mixed into the
high--redshift IGM.

To date, the reionization history has been constrained by observations
of Lyman series absorption spectra of $z\sim 6$ quasars (see
\citealt{fanreview} for a recent review)  by GRB observations at high redshifts \citep{totani06}, by
observations of Lyman-$\alpha$ selected galaxies 
\citep{kashikawa,mcquinn,dijkstra,malhotra} and  by CMB polarization
anisotropies \citep{kogutetal03,Pageetal07,spergeletal07}.  The
present constraints, however, are consistent with a wide range of
scenarios, and considerable theoretical and experimental effort is
devoted to developing new probes of the high-redshift universe.  A
promising possibility is to use the redshifted 21cm hyperfine line of
HI (see \citet{Furlanettoohreview} and references therein).  Metal
enrichment of the high--redshift IGM may be another useful tracer of
the reionization process. For example \citet{bhms} and \citet{CHVJ}
considered the elastic resonant scattering of CMB photons by
intergalactic metals. Recent studies have focused, in particular, on
the detectability of neutral oxygen (OI; \citealt{oh2002,bhms}), as
this element is thought to be produced in abundance by the first stars
\citep{hegerwoosley2002}.  Scattering of UV photons by OI, and the
corresponding absorption features in the spectra of quasars -- the OI
forest-- was proposed as a possible observable by \citet{oh2002}, and
may have recently been detected at $z\sim 6$ \citep{BeckerOforest06}.
Since OI and HI are in charge exchange equilibrium, oxygen is likely
to be highly ionized in regions where hydrogen is actively ionized.
However, the recombination time for oxygen is shorter than the Hubble
time, and it can be neutral even in ``fossil'' HII regions where
hydrogen had been ionized, but where short--lived ionizing sources
have turned off, allowing the region to recombine \citep{oh2002}.
\citet{ohhaiman03} show that the filling factor of such fossil HII
regions can be large ($\gsim 50\%$) prior to reionization.

In a previous paper (\citealt{HHJV}; hereafter Paper I), we showed
that the $63.2 \mu$m fine structure line of neutral OI can be pumped
by the $\sim 1300$\AA\ soft UV background, via the Balmer $\alpha$
line of OI. This is analogous to the Wouthuysen--Field effect for
exciting the 21cm line of cosmic HI. In Paper I, we found that OI at
redshift $z$ should be seen in emission at $(1+z)63.2\mu$m, and for
$7<z<10$ it would produce a mean spectral distortion of the CMB with a
$y$--parameter of $y\equiv \Delta I_{\nu} /B_{\nu}[T_{CMB}]=(10^{-9} - 3\times10^{-8}) (Z/10^{-3}{\rm
  Z_{\odot}}) (I_{UV}/10)$, where $Z$ is the mean metallicity of the
IGM and $I_{UV}$ is the UV background intensity at $1300$\AA\, in
units of $10^{-21}$ erg/s/Hz/cm$^2$/sr. In principle, this distortion
could be detectable through a precise future measurement of the CMB
spectrum (see \citealt{fixsenmather02} for the prospects of such
measurements), and would then open the possibility of performing
tomography of the metal distribution.  In combination with HI 21cm
studies, it could yield direct measurements of the abundance and
spatial distribution of metals in the high--redshift IGM.

Since oxygen pollution at high-redshift is likely associated with most
over-dense regions in the IGM, hosting the first star forming activity,
there must be spatial fluctuations in the OI abundance, causing
fluctuations in the corresponding $y$ distortion.  In this paper, we
consider the clustering properties of this signal. In particular, we
use toy--models to describe the metal distribution, and compute the
two--dimensional angular power spectrum of the CMB intensity. This would
be appropriate for an instrument with a single (or a few discrete)
frequency bands that probes the metals in a single (or a few discrete)
narrow redshift bins. We find that the clustering signal induced by
the inhomogeneous OI pumping is small, but possibly detectable with
forthcoming instruments, e.g., with deep integrations (for 2-3 months)
with the full ALMA array. A detection of the metal enrichment during
the Dark Ages would be complementary to HI 21cm studies, and would
provide clues about the distribution of metals in the IGM before
galaxy formation started in full.

The rest of this paper is organized as follows.
In \S~\ref{sec:pumping}, we discuss the basics of the OI pumping process.
In \S~\ref{sec:CMB}, we outline the computation of the distortion
induced by OI pumping along a given line of sight in the CMB.
In \S~\ref{sec:clustering}, we discuss our method to compute the
angular power spectrum of the distortion, with the basic assumption
that metals trace the distribution of collapsed halos.
In \S~\ref{sec:bubbles}, we introduce toy models to describe the
spatial distribution of metals, with the basic assumption that metals
cluster around dark matter halos. 
In \S~\ref{sec:results},  we present our main result on the angular
power spectrum, and discuss its dependence on the basic model
parameters, as well as its detectability.
Finally, in \S~\ref{sec:conclude}, we summarize the implications of
this work and offer our conclusions. Throughout this paper, we adopt a
set of ``concordance'' cosmological parameters for a flat universe,
$\Omega_m=0.29$, $\Omega_{\Lambda}=0.71$, $\Omega_b=0.047$, $h=0.72$,
with a power spectrum normalization $\sigma_{8}=0.75$ and slope
$n=0.99$.

\section{The Balmer-$\alpha$ pumping of OI}
\label{sec:pumping}

In Paper I, we computed the distortion in the CMB induced by neutral
oxygen in environments where UV radiation is pumping the fine
structure 63.2$\mu$m M1 transition (between the $n=2$ electronic
states $ ^3P_2$ and $ ^3P_1$) via the Balmer-$\alpha$ line at $\sim
1302$\AA\, connecting these two states with the excited $n=3$
electronic state $ ^3S_1$. In what follows, we shall denote the $
^3P_2$ and $ ^3P_1$ states as "0" and "1" respectively, and the
excited state $ ^3S_1$ as "2". Energy and frequency differences
between levels will be denoted as $E_{ji}$, $\nu_{ji}$, with
$i,j=0,1,2$ and $j>i$.

In Paper I, we showed that this process of {\it Balmer $\alpha$
  pumping} modifies the occupation of the levels $0$ and $1$ and
therefore introduces a shift in the spin temperature $T_S$, so that it
slightly departs from the CMB temperature, $T_{CMB}$.  Since at
reionization the UV background flux at $\nu_{10}$ is smaller than at
$\nu_{21}$, $T_S$ will be above $T_{CMB}$, producing an excess of 63.2
$\mu$m photons. Each of these excess photons is generated by the
sequence of three radiative transitions $0\rightarrow 2 \rightarrow 1
\rightarrow 0$, i.e., a photon of frequency $\nu_{20}$ is broken into
two photons of frequencies $\nu_{21}$ and $\nu_{10}$.

\begin{figure}
\plotone{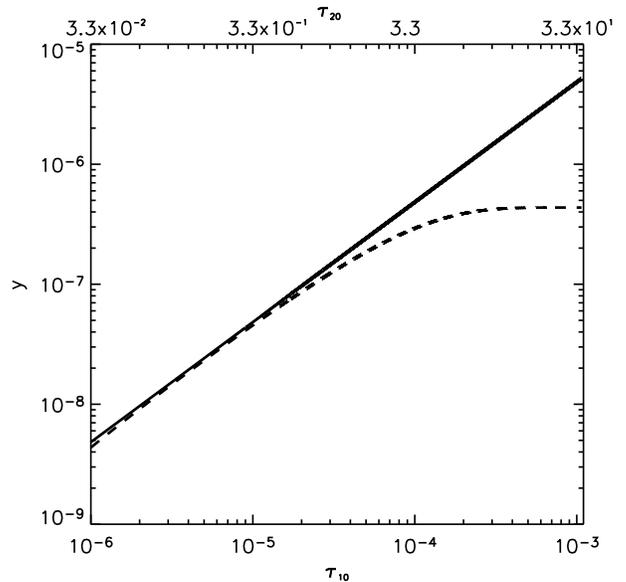}
\caption{The distortion parameter $y$ induced by the Balmer-$\alpha$
  pumping of the OI $63.2$ $\mu$m transition is shown as a function of
  the optical depths $\tau_{10}$ and $\tau_{20}$ at redshift
  $z=7$. The optically thin solution presented in Paper I is shown as
  the solid curve, accurately describing the system at low and
  intermediate $\tau_{10}$. When the 2$\leftrightarrow$0 transition
  becomes optically thick ($\tau_{20}\sim 1$, corresponding to
  $\tau_{10}\sim 3\times 10^{-5}$), the results are obtained by
  tracking the photon and level populations as dictated by
  equation~(\ref{eq:eqs1}), and shown by the dashed curve. Note that at low $\tau_{10}$ there is a 
  slight mismatch between the solid and the dashed lines, caused by numerical error due to the finite size
  of our integration step.}
\label{fig:s1}
\end{figure}

In Paper I, we were interested in the average distortion over the full
sky, and considered the optically thin limit for the Balmer-$\alpha$
transition.  While this is justified if the OI is homogeneously
distributed (in which case, $\tau_{20}\sim
0.01[Z/(10^{-2.5}Z_{\odot})]$ and $\tau_{21} \ll 1$), here we will
consider metal distributions that can be highly inhomogeneous.  The
metal--enriched bubbles can potentially contain a large concentration
of metals and become optically thick at $\nu_{20}$. For this reason,
we consider the case $\tau_{20}\gsim 1$ and self--consistently solve
the equations of radiative transfer to follow the UV intensity as the
background radiation penetrates the OI--rich bubble.

The equations governing the evolution of the relative populations
$n_i$ (with $i=0,1$) and the number of photons at the frequencies
$n_{\nu}^{2j}$, $j=1,2$ are:
\begin{eqnarray}
\nonumber
\frac{dn_0}{dt} & = & -n_0 \; \biggl[ \frac{g_1}{g_0}  A_{10} 
\frac{c^2I_{\nu}^{10}}{2h\nu_{10}^3}  \; + 
\;  P_{01}\biggr] \; + \; \\
 & & \phantom{xxxxxx}
              n_1\; A_{10}\; \biggl[   1 + \frac{c^2I_{\nu}^{10}}{2h\nu_{10}^3} \; +\; \frac{P_{10}}{A_{10}}\biggr] ,\; \\
\frac{dn_{\nu}^{20}}{dt} & = & c\; \frac{\lambda_{20}^2}{8\pi} \Psi_{\nu} \biggl[ -P_{01} n_0 \; + \; P_{10} n_1\biggr],
\label{eq:eqs1}
\end{eqnarray}
where $\Psi_{\nu}$ denotes the line profile, assumed to have a width
corresponding to a thermally broadened line with $T_{OI} \sim$ 20 eV  (corresponding
to a relative thermal width of $\sim 4\times 10^{-5}$).
For simplicity, we assume top-hat shapes for the line profiles, and we
include the effect of cosmological redshift by adopting an appropriate
effective length for the line-of-sight integral (discussed below).
In addition, atom and photon number conservation imply that $dn_1 / dt
= - dn_0 / dt$ and $d n_{\nu}^{21} / dt = - d n_{\nu}^{20} / dt$.

As initial conditions, we chose $T_S = T_{CMB}$, $I_{\nu}^{21} =
2\times 10^{-20}$ erg cm$^{-2}$ s$^{-1}$ Hz$^{-1}$ sr$^{-1}$, and
$I_{\nu}^{20} / I_{\nu}^{21} \simeq 1 - \beta$, with $\beta = 0.02$
as in Paper I (our results are largely insensitive to the choice
  of this parameter, as shown in Paper I).  Figure~\ref{fig:s1} shows
the result of evolving this system of equations at $z=7$. At low
$\tau_{10}$, the results (dashed curve) coincide with the optically
thin steady state solution of Paper I (solid curve) within a few
percent level corresponding to innacuracies in the numerical
integration. At larger values of $\tau_{10} \gsim 10^{-4}$, the
Balmer-$\alpha$ transition becomes optically thick and the number of
available pumping UV photons decreases. As a result, once these
photons are exhausted, the distortion $y$ flattens out at the maximum
value of $y \sim 3\times 10^{-7}$.

\section{The spectral distortions on the CMB}
\label{sec:CMB}

The total distortion introduced by OI in the CMB at a given observing
frequency $\nu_{obs}$ is
\begin{equation}
y = \int dr \; (j_{\nu} /B_{\nu}[T_0]) (r) \simeq  \frac{j_{\nu} }{B_{\nu}[T_0]}(\nu_{obs}) \Delta r,
\label{eq:ydef}
\end{equation}
where $j_{\nu}$ is an effective emissivity which is non zero only at
the redshift satisfying $1 + z_s \approx \nu_{10} / \nu_{obs}$ (see
Paper I for more details).  For a uniform OI distribution, and
assuming that the OI atoms follow the Hubble expansion, the effective
length $\Delta r$ along which this emissivity is not zero is
$cH^{-1}(z_s) \; (\Delta \nu_{th} / \nu_{10})$, where $\Delta
\nu_{th}$ is the width of the line.  However, if oxygen is clumped
(for example, confined within bubbles) then the effective length
$\Delta r$ may vary. If the physical size $L$ of the bubble is small,
$L\lsim cH^{-1}(z_s) \; (\Delta \nu_{th} / \nu_{10})$, then $\Delta r
\approx L$.  On the other hand, if $L\gsim cH^{-1}(z_s) \; (\Delta
\nu_{th} / \nu_{10})$ {\em and} the bubble has split from the Hubble
flow, then also $\Delta r = L$. But if the bubble is comoving with the
Hubble flow, then $\Delta r = cH^{-1}(z_s) \; (\Delta \nu_{th} /
\nu_{10})$,  (this length, at $z=7$, corresponds to $\sim 12$ Kpc). 
Note that, unless otherwise stated, distances will always be in physical units (and not
comoving). In this work, we shall assume that bubbles are not
gravitationally bound, and for simplicity, we further assume that they
expand with the Hubble flow, so that $\Delta r = \min{\left(L,
    \;cH^{-1}(z_s) \; \Delta \nu_{th} / \nu_{10}\right)}$. Hence, in
this case,

\begin{equation}
y \simeq  \frac{j_{\nu}} {B_{\nu}[T_0]} \min{\left(L, \;cH^{-1}(z_s)\; \frac{\Delta \nu_{th}} { \nu_{10}} \right)}.
\label{eq:yb}
\end{equation}

Depending on the angular and spectral resolution of a specific
instrument, the $y$--distortion along a given line--of--sight, and at
a given frequency, may have to be convolved with the respective
response functions (clearly, such smoothing can reduce the
fluctuations).  In what follows, we adopt a model instrumental beam
(or PSF) response function $V_{psf}(\vnh )$ and a frequency response
function $\Phi(\nu )$. Both will be taken to have accurately
determined Gaussian shapes, so that $\int d\vnh V_{psf}(\vnh ) = \int
d\nu \Phi(\nu ) = 1$, where $\vnh$ denotes the unit vector.

We assume generically that each OI--bubble is associated with a halo
of some mass $M$, with an abundance given by the Sheth-Tormen (ST,
\citealt{st}) mass function $dn/dM$, and that the OI atoms are
distributed in a bubble around the halo with a spherically symmetric
density profile $W_b(r)$; in this case, a second convolution, over the
bubble density profile, is necessary.

If we neglect the peculiar velocities and internal motions of the
bubbles, the effective distortion along the line of sight in direction
$\vnh$, and at frequency $\nu_{\rm obs}$, reads
\[
y_{eff} (\nu_{\rm obs},\vnh )= \int_{0}^{\infty} \int_{PSF}d\nu d\vnh
\Phi(\nu)  V_{psf}(\vnh ) \int_{r_{s}-\Delta r /2}^{r_{s}+\Delta r/2}
dr
\]
\begin{equation}
\times\int\int d\vy \; dM\; \frac{dn}{dM}(\vy,M) W_b(\vy-\vrv) \frac{j_{\nu}}{B_{\nu}[T_0]}
\label{eq:tau_oi_eff}
\end{equation}
where $\vrv = r\vnh$. The distance $r$ is centered at the resonant
value $r_{s}\equiv r[z_{s}]$ and $\Delta r$ corresponds to the
effective length defined above.  The integral over $M$ counts OI
bubbles around halos of various masses (above some minimum halo mass
$M_{min}$, as discussed further in \S~\ref{sec:bubbles} below),
whereas the integrals over $\nu$ and $\vnh$ describe the cosmological
volume probed by the PSF and the spectral response of the detector.
Let us change variables from frequencies and angles to spatial
coordinates, and define the three--dimensional instrumental PSF,
${\cal B}(\vx-\vrv_{s} )\equiv \Phi(\nu [x])
(\nu_{obs}/r_{s}^2)(dz/dr(z))V_{psf}(\vnh)$\footnote{The integral
  $\int d\vx {\cal B}(\vx -\vrv_{s}) = 1$ is normalized to unity in
  the volume centered at $\vrv_{s}$.}.  Further defining
$\hat{y}(M,z)\equiv (j_{\nu} /B_{\nu}[T_0]) \Delta r$, the distortion
generated by a bubble around a single halo of mass $M$ at redshift
$z$, we can re--write equation~(\ref{eq:tau_oi_eff}) more
transparently as
\begin{equation}
y_{eff} (\nu_{\rm obs},\vnh )\approx  \;\int \;dM\;\left[{\cal B} \star
  \left(\frac{dn}{dM}\star W_b\right)\right] \hat{y}.
\label{eq:tau_oi_eff3}
\end{equation}
where  $\star$ denotes convolution. In what follows, the twice
convolved halo number density will be denoted by $\tilde{n}$, i.e., 
$\tilde{n} (M,\vrv_{s})\equiv \left[{\cal B} \star \left( dn/dM\star W_b
\right)\right](M,\vrv_{s}) $

Finally, the physical collision between two metal-rich bubbles will
lead to a complicated thermo-dynamical interaction.  Rather than
modeling this process, the above prescription assumes a linear
addition of the OI density (or $y$--distortion) from two physically
overlapping metal bubbles.  This simple assumption at least captures
the enhancement of the metallicity in the overlap regions.  We note
that this is different from the case of merging HII bubbles during
reionization, when mergers conserve volume and result in the expansion
of the joint bubble.

\section{The Effect of Clustering}
\label{sec:clustering}

We next compute the second order moments (correlation function and
power spectrum) of the distortion field generated by OI at 63.2$\mu$m
during reionization.  Throughout this discussion, we will consider a
fixed redshift, and suppress the frequency--dependence of $y_{eff}(\vnh
)$. The angular correlation function can be written as
\[
\langle y_{eff}(\vnh_1) y_{eff} (\vnh_2)\rangle = \; \int \;dM_1 dM_2\; 
\langle \tilde{n}_1 \tilde{n}_2 \rangle
{\hat y}_1{\hat y}_2 
\]
\begin{equation}
=\int \;dM_1 dM_2\; 
\bar{\tilde{n}}_1 \bar{\tilde{n}}_2
{\hat y}_1{\hat y}_2\; \biggl(1 + {\tilde \xi}_{hh}(\vrv_1,M_1,\vrv_2,M_2) \biggr).
\label{eq:cf1}
\end{equation}
In this equation, $\bar{\tilde{ n}}$ denotes {\em average} ${\tilde
n}$.  The correlation function ${\tilde \xi}_{hh} (\vrv_1, M_1,
\vrv_2, M_2) $ corresponds to the halo-halo correlation function
convolved with {\em both} the window function of the profile of the OI
distribution ($W_b$) {\em and} the window function of the experiment
(${\cal B}$).  Keeping in mind that the Fourier counterpart of
${\xi}_{hh}$ is the matter power spectrum times the square of the bias
factor $b(M,z)$ computed in \citet{shethtormen99},
equation~(\ref{eq:cf1}) becomes
\[
\langle y_{eff}(\vnh_1) y_{eff} (\vnh_2)\rangle =
\int \;dM_1 dM_2\; 
\bar{\tilde{n}}_1 \bar{\tilde{n}}_2
{\hat y}_1{\hat y}_2\; b_1 b_2 \; \times
\]
\begin{equation}
\int \frac{d\vk}{(2\pi)^3} \; P_m(k,z_{rs}) \left| {\cal B}_{\vk}
\right|^2\left| W_{b,\vk}\right|^2 \exp{(-i\vk(\vrv_1-\vrv_2))},
\label{eq:cf2}
\end{equation}
with $P_m(k,z_{s})$ the matter power spectrum at redshift $z_s$, (note
that we have dropped a constant --$k=0$-- term and hence we are looking at the  {\em departure} of the correlation function from its mean value) and $W_{b,\vk}$ and
${\cal B}_{\vk}$ the Fourier counterparts of $W_b$ and ${\cal B}$
respectively.  Since we are assuming that every halo is producing a
bubble, and the dominant signal will be due to the clustering of
different bubbles, we can neglect non-linear corrections to the matter
power spectrum. Such non--linear corrections would boost the
clustering signals we predict below, but only on comoving scales of $\lsim 0.1$
Mpc at $z\approx 6$ (e.g. \citealt{iliev03}), which is well below the
scale at which the OI clustering signal peaks.  The integral on $\vk$
can be split into a transverse ($\vk_{\perp}$) and a parallel ($k_z$)
component along the line of sight. We assume that both ${\cal
B}_{\vk}, W_{b,\vk}$ can be factorised as ${\cal B}_{\vk} = ({\cal
B}_{\vk_{\perp}}\;{\cal B}_{k_z}) $ and $W_{b,\vk} =
(W_{b,\vk_{\perp}} W_{b,k_z})$. For a Gaussian instrumental response,
we can write ${\cal B}_{\vk_{\perp}} = \exp{-[k_{\perp}^2
(r_{s}\sigma_{b})^2/2]}$ and ${\cal B}_{k_z} = \exp{-[k_z^2
((dr(z)/dz)\sigma_{\nu}/\nu_{obs})^2/2]}$, with $\sigma_{b}$ related
to the width of the angular PSF and $\sigma_{\nu}$ to the width of
$\Phi(\nu )$. For simplicity, the bubble is taken to have a Gaussian
profile and a volume equal to $(\sqrt{2\pi}L)^3$.  After this
decomposition, the correlation function reads
\[
\langle y_{eff}(\vnh_1) y_{eff} (\vnh_2)\rangle = 
\phantom{xxxxxxxxxxx}
\]
\[ \int
\frac{d\vk_{\perp}}{(2\pi)^2}\;\exp{(-i\vk_{\perp}(\vrv_1-\vrv_2)_{\perp})}
\int dM_1 dM_2\; \bar{\tilde{n}}_1 \bar{\tilde{n}}_2
 {\hat y}_1{\hat y}_2\; b_1 b_2 
\]
\begin{equation}
\times \left| {\cal B}_{\vk_{\perp}} W_{b,\vk_{\perp}} \right|^2 \int \frac{d k_z}{2\pi}
\;\left| {\cal B}_{k_z} W_{b,k_z}\right|^2 P_m(k_{\perp},k_z, z_{rs}).
\label{eq:cf3}
\end{equation}
Finally, by noting that  in the flat sky approximation $\vk_{\perp}
\simeq \vl / r_{s}$,  we obtain the angular power spectrum:
\[ C_l = 
\frac{1}{r_{s}^2}
\int\int \;dM_1 dM_2\; 
\bar{\tilde{n}}_1 \bar{\tilde{n}}_2
{\hat y}_1{\hat y}_2\; b_1 b_2 \left | {\cal B}_{\vl/r_{s}} W_{b,\vl/r_s}\right|^2\; 
\]
\begin{equation}
 \int \frac{d k_z}{2\pi}
\;\left| {\cal B}_{k_z} W_{b,k_z}\right|^2 P_m(\vl/r_{s},k_z, z_{s}).
\label{eq:cl1}
\end{equation}

Let us briefly examine the behaviour of the $C_l$'s. We first recall
that the Fourier window function of a bubble is proportional to the
bubble volume ($\propto L^3$), and that the oxygen number density in
bubbles is proportional to $\bar{n}_{OI} / ({\bar N}L^3)$, with ${\bar
N}$ the average bubble number density, $\bar{n}_{OI}$ the global mean
oxygen number density at a given redshift, and $L$ the typical bubble
size. We also note that the bias factor $b$ almost cancels the
redshift dependence of the growth factor of perturbations (e.g.,
\citealt{ohcoorayk}), so the scaling of the maximum of $l^2C_l$ is
$(l^2C_l)_{max} \propto {\cal G}^2 \; \bar{n}_{OI}^2 \; (\Delta r)^2$,
with ${\cal G}$ a frequency-dependent function accounting for the
efficiency of the pumping process and $\Delta r$ the effective length
in equation~(\ref{eq:ydef}).  ${\cal G}$ dominates the overall
redshift dependence of the power spectrum amplitude.  If $L$ is
smaller than $cH^{-1}(z_s)\; \Delta \nu_{th} / \nu_{10}$, then
$(l^2C_l)_{max} \propto {\cal G}^2 \; \bar{n}_{OI}^2 \; L^2$, but
otherwise $(l^2C_l)_{max} \propto {\cal G}^2 \; \bar{n}_{OI}^2 $,
i.e., the band power spectrum is independent of the bubble size
(except that the bubble Fourier window function suppresses power at
scales smaller than the bubble size). In either case, note that $(l^2C_l)_{max} $ is proportional
to the square of the oxygen metallicity. However, if the OI abundance
within the bubbles is so high that they become optically thick at
$\nu_{20}$, then $y$ will reach its plateau, $y\sim {\cal G}$ and
$(l^2C_l)_{max} \propto {\cal G}^2 \; ({\bar N} L^2)^2$, i.e.,
$(l^2C_l)_{max} $ will be proportional to 
the square of the number
density of bubbles times their cross-sectional area,
assuming that the
frequency resolution of the instrument corresponds to a radial
distance that does not exceed the typical bubble size. Otherwise the scaling would be
$(l^2C_l)_{max} \propto {\cal G}^2 \; ({\bar N} L^3)^2$.
  The discrete nature
of the source distribution means that in addition to the correlation
term computed above, there will be a Poisson term. The Poisson
contribution can be computed by considering the limiting case of
equation~(\ref{eq:cl1}) for a random source distribution and an
experiment of infinite angular and spectral resolution,
\begin{equation}
C_l \simeq y^2 (\bar{N}L^3) \Delta \Omega_b
\label{eq:cl_p}
\end{equation}
with $y$ the average bubble distortion, $\bar{N}L^3$ the volume
fraction occupied by bubbles and $\Delta \Omega_b$ their typical
angular size.  In practice, we find that the shot noise contribution
is nearly always sub-dominant in the results we present below.

\section{Toy Model for Metal Distribution}
\label{sec:bubbles}

\begin{figure}
\plotone{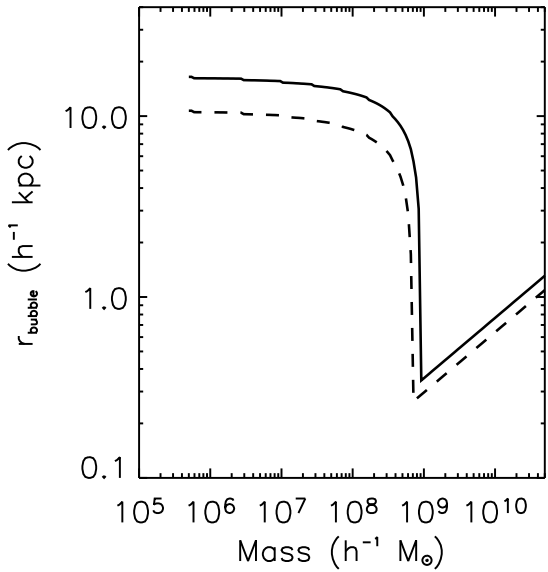}
\caption{The radii of metal bubbles around halos of mass $M$ produced
  in our fiducial toy--model, assuming metals expand over the typical
  age of the halo at a constant speed of $(v_{b}^2 - v_{esc}^2)^{1/2}$.
  Here $v_b$ is a wind velocity equal to $150$km/s, and $v_{esc}$ is
  the escape velocity from the halo. The solid curve shows the radii
  at $z=6.5$; the dashed curve at $z=7.5$. The sharp feature at $M
  \approx 10^9~{\rm M_\odot}$ corresponds to the limit where metals
  are trapped inside the halo potential well ($v_b \approx v_{esc}$).
  The bubble radius for more massive halos is assumed to be a fixed
  fraction (here taken to be 10\%) of the virial radius, and so
  it scales as $\propto M^{1/3}$.}
\label{fig:LvsM}
\end{figure}

To explore the detectability of the angular power spectrum of the
OI--induced CMB distortion, we need a model for the spatial
distribution of metals in the IGM.  Following the discussion in
\S~\ref{sec:CMB} and \S~\ref{sec:clustering} above, this means we need
to specify the profile $W_b(r)$ around a halo of mass $M$ at each
redshift $z$.

Our basic simplifying assumption is that metals are confined in
bubbles that had previously expanded into the IGM, but that these
bubbles have settled to follow the Hubble flow.  We assume that when a
halo is formed, a metal--polluting wind is launched from the galaxy at
the center of the halo, producing a bubble that subsequently expands
at a constant velocity for the age of the halo. In principle, halos of
the same mass that exist at a fixed redshift can have a distribution
of ages.  For simplicity, however, we assign a fixed age to halos of
mass $M$ at redshift $z$, corresponding to the average redshift $z_f$
at which such halos assembled 65\% of their mass (see, e.g., equation
2.26 in \citealt{lc93}).  At redshift $z$, the radius of the bubble
around a halo of mass $M$ is therefore given by
\begin{equation}
L(M,z) = 0.1 r_{vir}(M,z_f) + (v_b^2 - v_{esc}^2)^{1/2}
\left[t(z)-t(z_f)\right],
\label{eq:bubblesize}
\end{equation}
with $r_{vir}$ the virial radius of the parent halo, and $t(z)$ the
cosmic time at redshift $z$.  The first term on the right hand size
represents a rough estimate for the size of the galaxy in the halo
(i.e., the region producing the metals).  The velocity $v_b$, taken to
be a constant in the range of $50-1500~{\rm km~s^{-1}}$, represents
the wind velocity ``in vacuum''.  We then subtract $v_{esc}$,
corresponding to the escape velocity from the parent halo at a
distance of 0.1 $r_{vir}(M,z_f)$, from this wind velocity, to take
into account the fact that metals loose energy as they travel out of
the gravitational potential well.   Since we assume that metals have no peculiar 
velocities, the bubbles start comoving with the Hubble flow by the time they are observed.
Equation~(\ref{eq:bubblesize})
assumes further that star--formation takes place instantly when the
halo forms.  In Figure~\ref{fig:LvsM}, we show the bubble sizes
predicted under these assumptions.

For simplicity, we normalize the total amount of oxygen mass in all
bubbles at a given redshift by specifying the global average
metallicity of the IGM at each redshift, i.e.,
\begin{equation}
Z_{av} (z)= \int_{M_{min}} dM \frac{dn}{dM}(M,z) Z(M,z) {\cal V}_b (M,z),
\label{eq:zav}
\end{equation}
with $Z(M,z)$ the metallicity within a bubble at redshift $z$
originated in a halo of mass $M$ and with volume ${\cal V}_b = 4\pi/3
L^3$, and $dn/dM$ is the halo mass function.  For halos with $v_b >
v_{esc}$, we further assume that, at a given redshift, the oxygen
abundance within the bubble is proportional to the mass of the parent
halo.  Provided that the bubble radii are practically constant in this mass range, 
this is roughly equivalent to assuming that the star formation
rate scales linearly with halo mass, and is constant throughout the
age of the halo. For halos with $v_b < v_{esc}$, we assume that the
oxygen density is constant (since the metals produced in these halos
are assumed to be trapped in a volume ${\cal V}_b \propto M$);
however, these larger halos contribute negligibly to the clustering
signal below and our results are insensitive to this assumption.

\begin{figure*}
\plotancho{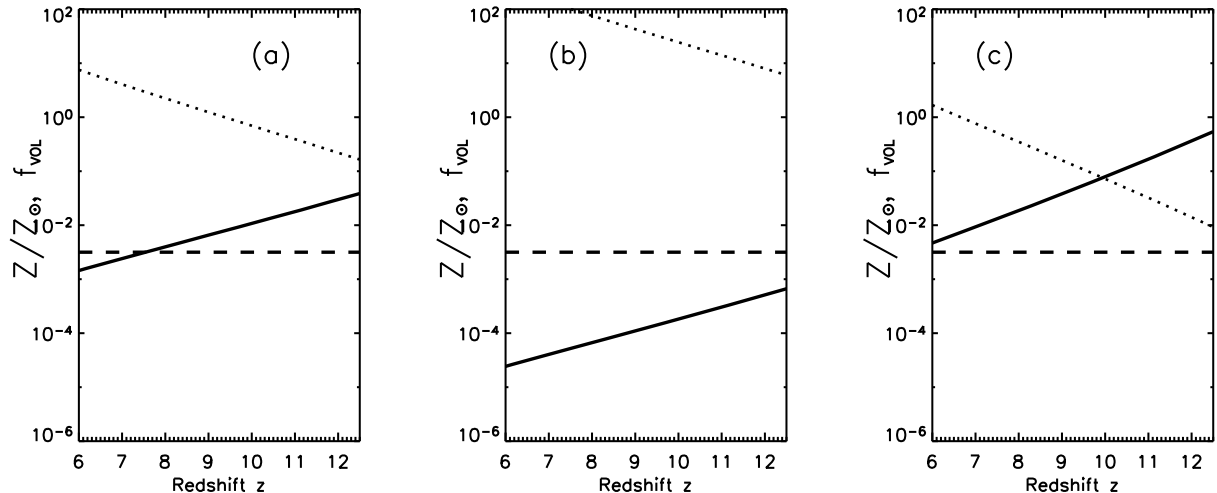}
\caption{Models of metal enrichment in the IGM at various redshifts.
  The global average metallicity $Z/Z_{\odot}$ is shown by the dashed
  curves, and the metallicity inside individual metal bubbles by the
  solid curves. Note that the global metallicity is renormalized at
  each redshift by hand to the same metallicity, and the curves should
  not be interpreted as evolutionary models.  The dotted curves show
  the volume--filling fraction occupied by bubbles (note that
  metal--rich bubbles can overlap and this fraction can exceed
  unity). Panel {\it (a)}: $M_{min} = 5\times 10^5$ M$_{\odot}$, $v_b
  = 60$ km/s and global average $Z/Z_{\odot} = 10^{-2.5}$ set to a
  constant in redshift. At $z \simeq 6.5$ the typical bubble size is
  $L\sim 22$ kpc, and bubbles have been generated in halos of typical
  mass $M \sim 4\times 10^6$ M$_{\odot}$. Panel {\it (b)}: same as
  panel (a) but $v_b = 150$ km/s, which corresponds, at $z\simeq 6.5$,
  to a typical bubble size of $L \sim 60$ kpc.  Panel {\it (c)} $v_b =
  150$ km/s (corresponding to $L \sim 50$ kpc at $z\simeq 6.5$) and
  $M_{min} = 5\times 10^7$ M$_{\odot}$.  }
\label{fig:zev}
\end{figure*}

In Figure~\ref{fig:zev}, we show the evolution of the global average
(dashed lines), and bubble-volume weighted (solid curves)
metallicities versus redshift. The dotted curves show the fraction of
the total volume contained inside the bubbles (note that metal--rich
bubbles can overlap and this fraction can exceed unity). In all
panels, the global average oxygen abundance is scaled to $10^{-2.5}
Z_{\odot}$ at each redshift. This value is not unrealistic, as there
are observational constraints on the cosmic metallicity indicating
that $Z/Z_{\odot} \sim 10^{-3}$ at $z>4$ \citep{schaye}, and oxygen
from first--generation stars is likely to be somewhat overabundant
relative to the solar value \citep{meynet}.  The highest redshift,
$z=12$, shown in the figures represents the earliest epoch when this
floor metallicity may have been established
(e.g. \citealt{haimanloeb97}). Note that since the global metallicity
is re--scaled at each redshift by hand, the curves in
Figure~\ref{fig:zev} should be interpreted as independent models of
the metal distribution at each redshift, and not as evolutionary
models.  In order to illustrate the metal distribution under different
assumptions, in panel (a), we assume $M_{min} = 5\times 10^{5}$
M$_{\odot}$ and $v_{b} = 50$ km/s; panel (b) is the same as (a),
except with $v_{b} = 150$ km/s, while in panel (c), we assume $v_b =
150$ km/s and $M_{min} = 5\times 10^{7}$ M$_{\odot}$. 

The enrichment scenarios considered in this work are comparable with those presented in previous studies. Indeed, in \citet{furlanettoloeb03} the typical bubbles sizes at $z\sim 6$ are a few tens of kpc, in agreement with our estimates, whereas, after looking at the simulations of \citet{finlator}, we have found similar enrichment levels in the relevant redshift ranges ($z\in[5,10]$). Only in \citet{scannapieco} the bubble filling factors tend to be slightly smaller than in our case, since, motivated by WMAP observations, we are considering 
early reionization models.

\section{Results and Discussion}
\label{sec:results}

\begin{figure*}
\plotancho{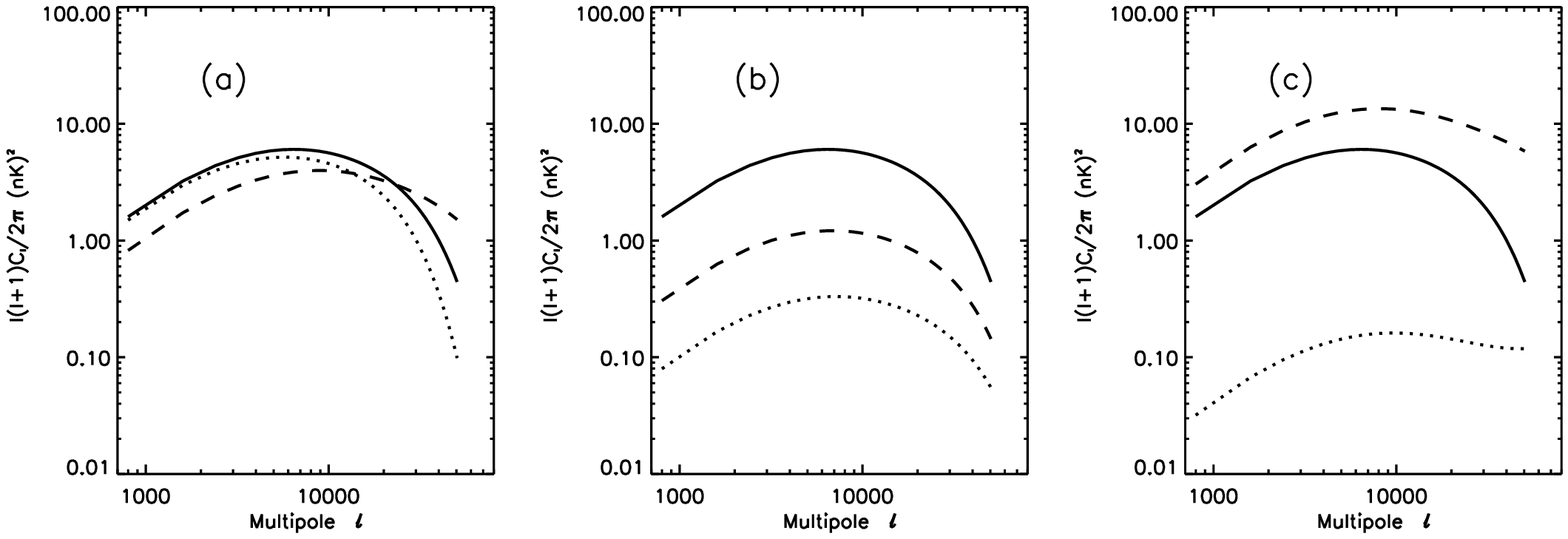}
\caption{The dependence of the angular fluctuation power, $l^2C_l$, on
  metal distribution model parameters and on redshift. In all three
  panels, we display our adopted fiducial model given by: $z=6.5$
  (observable at 630 GHz), $I_{\nu}^{20}=6\times 10^{-20}$erg
  cm$^{-2}$ s$^{-1}$, Hz$^{-1}$ sr$^{-1}$, $M_{min}=5\times
  10^{5}$M$_{\odot}$ and $L\simeq 60$ kpc.  Panel {\it (a)}:
  dependence of the signal on bubble size: $23, 60, 80$ kpc for the
  dashed, solid and dotted curve, respectively. For $L
  \simeq 23$ kpc we still have $L < cH^{-1}(z_s) \Delta
  \nu_{th}/\nu_{10}$ and $(l^2C_l)_{max} \propto {\cal G}^2 L^2$, but
  larger values of $L$ imply $L \gsim cH^{-1}(z_s) \Delta
  \nu_{th}/\nu_{10}$ and the scaling changes to $(l^2C_l)_{max}
  \propto {\cal G}^2$. The dotted line  ($L \simeq 80$ kpc) shows
  how power is suppressed in scales below the typical bubble
  size. Panel {\it (b)}: dependence of the signal on redshift: $z=
  6.5, 7, 7.5$ from top to bottom, corresponding to observing frequencies $\nu = 633, 593$ and $558$ GHz, respectively. 
  Panel {\it (c)}: dependence on the
  minimum halo mass: $M_{min} = 5\times 10^5$ (solid line), $ 5\times 10^8$ (dashed line), and
  $7.5\times 10^8$ M$_{\odot}$ (dotted line). The
  distortion $y$ becomes saturated for $M_{min} \;_{\sim}^{>} 5\times
  10^8$ M$_{\odot}$. For $5\times 10^5{\rm M_\odot}<M_{min}< 5\times
  10^8$ M$_{\odot}$, we have the approximate scaling $(l^2C_l)_{max}
  \propto {\cal G}^2$ that is slightly distorted by the presence of
  the bias factor enhancing the dashed curve over the solid
  one. For $M_{min} = 7.5\times 10^8 $ M$_{\odot}$ we are well in the
  optically thick regime ($C_l \propto ({\bar N} L^3)^2$) 
{\em and}
  the typical bubble sizes have shrinked dramatically (see
  Fig.~\ref{fig:LvsM}; this translates into a clear drop in the power
  (thinnest curve).  }
\label{fig:y}
\end{figure*}

The angular power spectrum (correlation term plus shot noise term) of
the CMB distortion produced by the OI pumping is shown in
Figure~\ref{fig:y}. We have adopted an angular resolution of one
arc-second, and a relative spectral resolution of $\Delta\nu/\nu=10^{-4}$. 
The angular resolution corresponds to $\ell \sim 7\times 10^5$ and
therefore should resolve all bubbles under consideration. The frequency
resolution corresponds at $z=6.5$  to roughly 60 kpc in physical units.
In all three panels, we consider variations from our adopted fiducial
model given by $z=6.5$ (observable at 630 GHz), $I_{\nu}^{20}=6\times
10^{-20}$erg cm$^{-2}$ s$^{-1}$ Hz$^{-1}$ sr$^{-1}$, $M_{min}=5\times
10^{5}$M$_{\odot}$, and $v_b=150$ km/s, corresponding to $L \simeq 60$
kpc at that redshift.  Panel {\it (a)} shows the dependence of the
signal on bubble expansion velocity/bubble size: 150 (solid curve),
60 (dashed curve) , and 200 (dotted curve) km/s, which correspond to $L \simeq$ 60,
23 and 80 kpc respectively. 
If $v_b = 60$ km/s ($L\simeq 23$ kpc), then $L < cH^{-1}(z_s) \Delta
\nu_{th}/\nu_{10}$, and the bubbles remain optically thin ($y$ has not
reached its maximum value), so that $(l^2C_l)_{max} \propto {\cal
  G}^2\;L^2$. On the other hand, for $v_b {\sim} 150 $ km/s ($L \simeq
$60 kpc), we are in the regime of $L \;^>_{\sim} cH^{-1}(z_s) \Delta
\nu_{th}/\nu_{10}$ and $(l^2C_l)_{max} \propto {\cal G}^2$. Therefore, if $v_b =
200$ km/s, the amplitude of the spectrum should not change noticeably. However, the bubble size ($L\simeq 80$ kpc) is slightly larger than in our fiducial model ($L\simeq 60$ kpc), and
when comparing both cases one can clearly see the effect of the bubble window function suppressing
power at small scales.
Therefore $(l^2C_l)_{max} \propto (v_b/ 150
\;\mbox{km/s})^2$ if $v_b < 150$ km/s. In panel {\it (b)}, we find a
strong dependence of the power on the redshift: for the range of
$z_s=6.5, 7, 7.5$ (from top to bottom), the amplitude of the power
spectrum varies by nearly two orders of magnitude. This steep
dependence arises because the efficiency of the pumping is
exponentially dependent on the difference between the CMB and the spin
temperatures, and this difference is reduced at higher $z$ (see Paper
I for more discussion).

Panel {\it (c)} in Figure~\ref{fig:y} shows the dependence on the
minimum size of the halo producing metal--bubbles, with $M_{min} =
5\times 10^5$ (solid line), $5\times 10^8$ (dashed line), and 
$7.5\times 10^8$ M$_{\odot}$ (dotted line). 
This roughly brackets the possible range
of minimum halo sizes, based on the presence or absence of ${\rm H_2}$
molecules (e.g. \citealt{haimanreesloeb97}).  At $M_{min} = 5\times
10^5$ M$_{\odot}$, $L \;\sim cH^{-1}(z_s) \Delta \nu_{th}/\nu_{10}$,
which would yield the scaling $(l^2C_l)_{max} \propto {\cal G}^2 $,
i.e., independent on $M_{min}$. However, the actual amplitude does
vary with mass in the range of $5\times 10^5$ M$_{\sun} < M_{min} <
7.5\times 10^8$ M$_{\sun}$ due to several reasons. First, increasing
the halo mass results in a larger bias factor (from $b\simeq 1.4$ to
$b\simeq 2.6$). Second, at large $M_{min}$, there are fewer bubbles
per unit volume, assumed to contain the same amount of metals, so the
bubbles become optically thick, and $y\sim {\cal G}$ and
$(l^2C_l)_{max} \propto {\cal G}^2 \; ({\bar N} L^3)^2$. But most importantly, for
$M_{min} > 5\times 10^8 {\rm M_\odot}$, the typical bubble radii (and
corresponding volumes) shrink considerably (c.f.Fig. \ref{fig:LvsM}),
and due to the scaling $C_l \propto ({\bar N} L^3)^2$ , this causes a
correspondingly steep drop in the power amplitude (dotted line in
Fig.(\ref{fig:y}c)).  We also note that for the largest minimum mass
considered, the shot noise contribution starts dominating at the
smallest scales.  We find that the amplitude of the signal
depends strongly on the parameters of the model. However, the toy model for the metal distribution was
introduced only to gain insight on the nature of the signal and should be regarded as a toy model.
Numerical simulations with radiative transfer should be used to make more quantitative predictions and
will be crucial in the interpretation of future observations.

The multipole corresponding to $(l^2C_l)_{max}$ in
Figure~\ref{fig:y} is related to the projected $k$ mode at which the
convolved matter power spectrum peaks. In principle, the typical bubble clustering length could also be
detected in the {\em radial} direction if the observing instrument had
enough spectral resolution: in such case, observations at slightly
different frequencies would give rise to {\em correlated} maps, since
they would be probing overlapping shells centered at similar
redshifts. The combination of maps obtained at different frequencies
could also improve the S/N ratio of the final detection.

Because at the frequencies of interest the signal will be dominated by
infrared emission from dusty galaxies, it is important to be able to
remove them from the observed map. One way this can be achieved is by
obtaining high angular resolution images of the sky at the frequencies
of interest, so that Olber's paradox is avoided and individual
galaxies can be clipped from the maps. The Atacama Large Millimeter
Array (ALMA) may be an excellent instrument for this purpose because
of its high sensitivity and angular resolution (of the order of
0.5"). The drawback is that the field of view of ALMA is small
($\lsim$ 1 arcmin). However, since our signal peaks at scales of
$\sim$ tens of arcseconds and decays slowly at smaller scales, this
small field of view should not a fundamental limitation. 
The signal could also be within reach of the sensitivity of forthcoming detectors (e.g., SCUBA2, \citep{holland}) and of planned single dish experiments such as CCAT\footnote{CCAT URL site: {\tt http://www.submm.org}}. However, current small scale CMB experiments, with a typical sensitivity 
of $\sim \mu$K at arcmin scales, are still rather far from imposing interesting constraints on this effect. 

\section{Conclusions}
\label{sec:conclude}

In this paper, we have computed the angular fluctuation of the effect
we presented in an earlier work (Paper I): the CMB spectral distortion
induced by OI $63.2$ $\mu$m fine transition, pumped at the epoch of
reionization by the UV background radiation via the OI Balmer $\alpha$
line.  Since the oxygen distribution in space is associated with the
most over-dense regions hosting the first star forming activity at the
end of the Dark Ages, there must be fluctuations in the angular
pattern of the OI-induced $y$ distortion.  We have considered the
clustering properties of the signal and computed its angular power
spectrum using a toy model for the metal distribution.  We find that
the predicted signal is small, but that for certain enrichment models
it could be detected using ultra-deep observations with a sensitive
instrument such as ALMA, SCUBA2 and CCAT.  Even in the event of a non-detection, 
future infrared observations
could place interesting limits on the metal enrichment of the universe
during the Dark Ages. This could open the possibility of measuring the
metal enrichment of the universe before it was re-ionized. OI
observations would also be complementary to HI 21cm measurements that
will map out in the near future the distribution of neutral H during
the epoch or re-ionization.

\acknowledgements
ZH acknowledges partial support by NASA grant NNG04GI88G. CHM, LV and RJ acknowledge support by  NASA grant ADP04-0093, and NSF grant PIRE-0507768. The authors thank Chris Carilli for useful discussions.

\clearpage

\end{document}